\documentclass[aps,pra,twocolumn,showpacs,superscriptaddress,floatfix]{revtex4-1}
\usepackage{graphicx,amsmath,amssymb}
\usepackage[usenames]{color}
\usepackage[dvipsnames]{xcolor}
\usepackage[unicode=true,pdfusetitle,
 bookmarks=false,
 breaklinks=false,pdfborder={0 0 1},backref=false,colorlinks,urlcolor=Black, linkcolor=blue,citecolor=blue]
           {hyperref}
\newcommand{\ZZ}{\mathbb{Z}}

\begin{document}

\title{Critical quantum metrology in a stabilized two-photon Rabi model}

\author{Zu-Jian Ying}
\email{yingzj@lzu.edu.cn}
\affiliation{School of Physical Science and Technology, Lanzhou University, Lanzhou 730000, China}
\affiliation{Key Laboratory for Quantum Theory and Applications of MoE, Lanzhou Center for Theoretical Physics, Lanzhou University, Lanzhou 730000, China}

\author{Hang-Hang Han}
\affiliation{School of Physical Science and Technology, Lanzhou University, Lanzhou 730000, China}
\affiliation{Key Laboratory for Quantum Theory and Applications of MoE, Lanzhou Center for Theoretical Physics, Lanzhou University, Lanzhou 730000, China}

\author{Bo-Jian Li}
\affiliation{School of Physical Science and Technology, Lanzhou University, Lanzhou 730000, China}
\affiliation{Key Laboratory for Quantum Theory and Applications of MoE, Lanzhou Center for Theoretical Physics, Lanzhou University, Lanzhou 730000, China}

\author{Simone Felicetti}
\email{simone.felicetti@cnr.it}
\affiliation{Institute for Complex Systems, National Research Council (ISC-CNR),  00185 Rome, Italy}
\affiliation{Physics Department, Sapienza University, 00185 Rome, Italy}

\author{Daniel Braak}
\email{daniel.braak@uni-a.de}
\affiliation{TP III and Center for Electronic Correlations and Magnetism, University of Augsburg,
86135 Augsburg, Germany}

\begin{abstract}

We investigate a generalized quantum Rabi model (QRM) with  two- and four-photon terms with respect to applications
for non-linear critical quantum metrology. In the introduced model, the spectral collapse occurring in
the standard two-photon QRM is stabilized by the presence of the quartic potential. The collapse is then transformed into a quantum phase transition, which occurs
in the low-frequency limit of the light mode, whose remnant at finite ratio between qubit and mode frequencies can be
applied to critically enhanced quantum metrology. We find that the four-photon term entails a much higher
measurement precision compared to the standard two-photon QRM. The mechanism behind the higher precision can be traced to the different behavior of the ground state wave function as the system is tuned through the transition. As the standard two-photon QRM, despite the absence of the spectral collapse, our model allows for a
 finite preparation time for the probe state (PTPS).
\end{abstract}
\pacs{ }
\maketitle

\section{Introduction}

Quantum metrology and sensing are based on the use of  quantum resources~\cite{Paris2009,RevModPhys_QSensing} to overcome the performance of any classical strategy. Critical quantum sensing (CQS) is by now a well established approach, which makes use the quantum properties spontaneously developed by critical systems in proximity of quantum phase transitions.
Indeed, it has been theoretically shown that a quantum-enhanced sensing precision can be achieved exploiting static~\cite{Zanardi2008,ivanov_adiabatic_2013,Bina2016,Lorenzo2017,Ivanov2020,invernizzi2008Optimal,Mirkhalaf2020,Niezgoda2021,DiFresco2022,DiFresco2024,Sahoo24} or dynamical~\cite{Tsang2013,macieszczak_dynamical_2016,Cabot24,zicari2024} properties of many-body systems in proximity of the critical point. In spite of the critical slowing down, the optimal precision scaling can be achieved not only with respect to the system probe size, but also with respect to protocol duration time~\cite{Rams2018}.
Furthermore, when using adaptive strategies~\cite{Montenegro2021,Salvia2023, PhysRevLett.133.120601}, CQS protocols are efficient also for global quantum sensing. The practical relevance of the CQS approach is confirmed by the first experimental implementations, carried out with Rydberg atoms~\cite{Ding2022}, nuclear magnetic resonance~\cite{Liu2021} and superconducting quantum circuits~\cite{Petrovnin_2024,beaulieu2024,yu2025experimental}.

Of particular interest for this work, CQS protocols can also be conceived~\cite{Garbe2020} considering finite-component phase transitions (FCPTs)~\cite{Ashhab2013,Ying2015,Hwang2015PRL,Hwang2016PRL,Irish2017,
LiuM2017PRL,Ying-2018-arxiv,Ying2020-nonlinear-bias,Liu2021AQT,Ying-2021-AQT,Ying-gapped-top,Ying-Stark-top,Ying-Spin-Winding,
Ying-JC-winding,Ying-Topo-JC-nonHermitian,Ying-Topo-JC-nonHermitian-Fisher,Ying-gC-by-QFI-2024, Grimaudo2022q2QPT,Grimaudo2023-Entropy,Grimaudo2024PRR,Zhu2024PRL,DeepStrong-JC-Huang-2024}, which are quantum criticalities formally obtained by applying a scaling limit on the system parameters~\cite{hwang_quantum_2015,Felicetti2020,LiuM2017PRL,Ying-2021-AQT,Ying-Stark-top}. FCPTs have been mainly studied considering quantum resonators with embedded atomic~\cite{Ashhab2013, Hwang2015PRL,Puebla2017,Peng2019,Zhu2020,Zhang24} or Kerr~\cite{Bartolo2016,Felicetti2020,Minganti2023a,Minganti2023b,LiuGang2023} nonlinearities, and they have been controllably implemented using atomic systems~\cite{cai2021observation},  polaritonic systems~\cite{DelteilNatMat19,Zejian2022} and superconducting quantum systems~\cite{Fink2017,Brooks2021,beaulieu2023observation,chen2023quantum,Sett24,jouanny2024}. FCPTs provide a tractable framework to analyze the performance of CQS protocols. For example, they have been used to demonstrate the constant-factor advantage of dynamical over statical protocols~\cite{Chu2021,Garbe2022}, to unveil the presence of apparent super-Heisenberg scalings~\cite{Gietka2022,Gietka2022-ProbeTime,Garbe2022a}, to analyze continuous-measurement schemes~\cite{Ilias2022,Yang2022}, and to make formal comparison with passive quantum sensing strategies~\cite{alushi2024optimality}.
This framework has been used to design CQS protocols which can be implemented with small-scale devices, considering systems based on parametric resonators~\cite{heugel2020_quantum,DiCandia2023,Rinaldi2021,Petrovnin_2024,Hotter24,choi2024observing,alushi_collective_2025}, trapped-ions~\cite{Ilias2023}, optomechanical~\cite{Bin2019,Tang2023} or magnomechanical~\cite{Wan2023} devices, spin defects~\cite{mih23multiparameter,mihailescu2024} and Rabi systems~\cite{Ying2022-Metrology,Xie2022,Gietka2023PRL-Squeezing,Zhu24,Ying-g2hz-QFI-2024,Ying-g1g2hz-QFI-2025}. \\

An interesting class of finite-component models is given by two-photon or quadratic couplings, where quantum emitters interact with bosonic modes exclusively via the exchange of excitation pairs. These models can be feasibly implemented with atomic~\cite{Felicetti2015-TwoPhotonProcess} and solid-state~\cite{Felicetti2018-mixed-TPP-SPP,Felicetti_Ultrastrong,Gautier2022,Ayyash2024,Wang2016,Sanchez2018} quantum technologies. Two-photon interactions lead to an interesting phenomenology~\cite{Ying-2018-arxiv,Cong2020,Ying2020-nonlinear-bias,Ma2020Nonlinear,Zou2020,Piccione2022} and they can induce quantum phase transitions in both extended~\cite{Garbe2020, Li2022, Li2024,shah2024} and finite-component models~\cite{Cui_2017,Ying2020-nonlinear-bias,Ying-2018-arxiv}. It has already been shown that two-photon couplings can be effectively deployed in critical quantum sensing protocols~\cite{Ying2022-Metrology,Ying-g2hz-QFI-2024,Ying-g1g2hz-QFI-2025}. Two-photon couplings are known to induce a spectral collapse, where the discrete spectrum collapses into a continuous energy band~\cite{Travenec2012,Duan2016,CongLei2019,Rico2020,Braak2023AnnPhys}. Although the onset of the collapse can in principle be observed~\cite{Felicetti2015-TwoPhotonProcess,Felicetti_Ultrastrong}, beyond the collapse point the model is unbounded from below and so becomes unstable.

Here, we introduce a generalized two-photon quantum Rabi model (QRM) that is stable in the whole parameter space. In particular, we include a quartic potential that  transforms the spectral collapse in a quantum phase transition. We derive analytical solutions for the system eigenspectrum in the low-frequency limit and characterize the system critical behavior. Finally, we assess the metrological power of this model by evaluating the quantum Fisher information over the ground state manifold. We find that the quartic potential not only stabilizes the model, but it also enhances the achievable precision in a critical quantum sensing protocol.

The paper is organized as follows. In
section~\ref{Sect-Model}, the non-linear QRM with a quartic regulator term ($A_4$ term) is introduced.
Section~\ref{Sect-Cure-ej-collapse} shows that the quartic term removes the ``spectral collapse'': The spectrum stays discrete for all parameter values.
Section~\ref{Sect-QPT-A4} studies the quantum phase transition induced by the $A_4$ term in the slow-mode limit and gives analytic expressions for the transition point.
Section~\ref{Sect-higher-QFI-A4} demonstrates a much enhanced QFI by the $A_4$ term, indicating higher measurement precision in quantum metrology as compared to the standard two-photon QRM.
Section~\ref{Sect-Mechanism-A4} clarifies the mechanism of higher sensitivity effected by the $A_4$ term.
Section~\ref{Sect-ProbeTime-A4} shows that the PTPS stays finite for experimentally realizable parameter values and
section~\ref{Sect-Conclusions-A4} presents concluding remarks.

\section{Model}
\label{Sect-Model}

The (extended) two-photon QRM is described by the following Hamiltonian~\cite{Ying-2018-arxiv,Ying2020-nonlinear-bias}
\begin{equation}
H_{{\rm T}}=\omega a^{\dagger }a+\frac{\Omega }{2}\hat{\sigma}_{x}+g_{2}\hat{%
\sigma}_{z}[(a^{\dagger })^{2}+a^{2}+\chi (2a^{\dagger }a+1)]  \label{H-T}
\end{equation}%
which contains a quadratic coupling between the bosonic mode with frequency
$\omega$, created (annihilated) by $a^{\dagger }$ ($a$), and a qubit
represented by the Pauli matrices $\hat{\sigma}_{x,y,z}$. The simple
two-photon QRM has $\chi =0$, corresponding to two-photon
absorption and emission processes. Including the Stark-like
term~\cite{Eckle-2017JPA} leads for $\chi =1$ to the form $\hat{\sigma}_{z}(a^{\dagger }+a)^{2}$ which is realizable in superconducting circuit systems~\cite{Felicetti2018-mixed-TPP-SPP}.
The $\ZZ_4$-symmetry $P_{4}=\hat{\sigma} _{x}e^{i\pi a^{\dagger }a/2}$ for $\chi =0$ is broken down to $\ZZ_2$ for $\chi \neq 0$~\cite{Ying-g2hz-QFI-2024}, with symmetry operator $P_{2}=e^{i\pi a^{\dagger }a}$~\cite{Ying-2021-AQT,Ying-JC-winding,Ying-g2hz-QFI-2024}.
The Stark-like term leads to a rescaling of the
critical coupling for spectral collapse \cite{Ying-2018-arxiv,Ying2020-nonlinear-bias}
\begin{equation}
g_{{\rm T}}=\frac{\omega }{2\left( 1+\chi \right) }.
\end{equation}%
In the following, we set $\chi =1$.

We augment now the Hamiltonian \eqref{H-T} by a term quartic in the boson operators, corresponding to four-photon processes,
\begin{equation}
H=\omega a^{\dagger }a+\frac{\Omega }{2}\hat{\sigma}_{x}+g_{2}\hat{\sigma}%
_{z}(a^{\dagger }+a)^{2}+A_{4}(a^{\dagger }+a)^{4},  \label{H}
\end{equation}
which can be realized in superconducting circuit systems~\cite{HangHang2024A4}.
This term does not couple to the qubit degree of freedom, it can therefore be called ``neutral'' with respect to the qubit. Nevertheless it has a profound effect on the spectrum and the dynamical behavior of model.
Here we have adopted the spin notation as in Ref.~\cite{Irish2014}, in which
the spin value in the $z$ direction, $\sigma _{z}=\pm $,
represents the two flux states in the flux-qubit circuit system\cite%
{flux-qubit-Mooij-1999}. This platform allows to realize the
ultra-strong~\cite{Ciuti2005EarlyUSC,Aji2009EarlyUSC,Diaz2019RevModPhy,Kockum2019NRP,Wallraff2004,Gunter2009, Niemczyk2010,Peropadre2010,FornDiaz2017,Forn-Diaz2010,Scalari2012,Xiang2013,Yoshihara2017NatPhys,Kockum2017,Ulstrong-JC-2,Ulstrong-JC-3-Adam-2019,PRX-Xie-Anistropy,Qin2024PhysRep}
and even deep-strong coupling regime~\cite{Yoshihara2017NatPhys,Bayer2017DeepStrong,Ulstrong-JC-1,DeepStrong-JC-Huang-2024},
with coupling strengths $g_2$ beyond $0.1\omega $ and $1.0\omega$.
In cavity QED systems $\Omega$ denotes the level splitting of the qubit and the model is usually written in a basis where the qubit is diagonal, corresponding to exchange of $\sigma_x$ and $\sigma_z$.

\begin{figure*}[t]
\includegraphics[width=2\columnwidth]{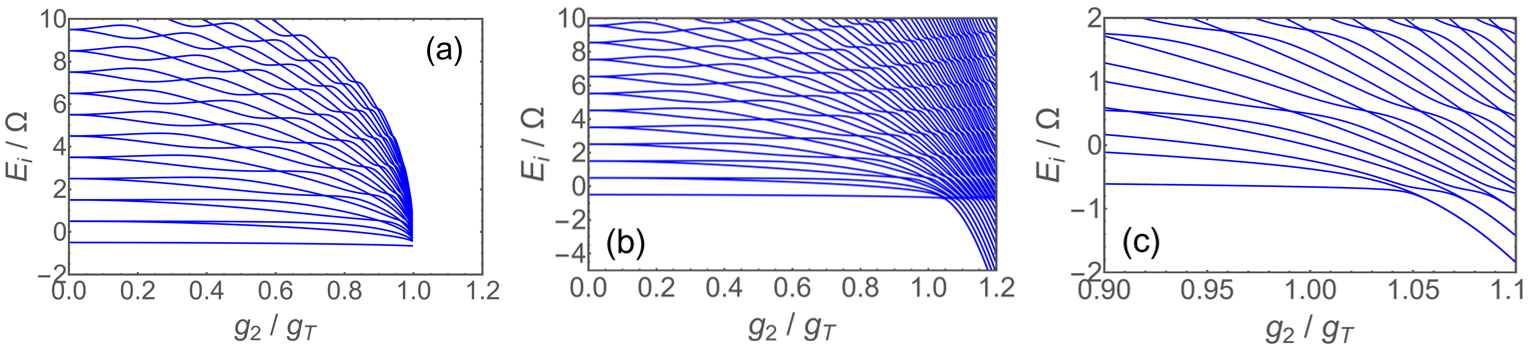}
\caption{The spectral collapse phenomenon. Energy spectrum $E_i$ at $\omega=1.0\Omega$:
(a) $A_4 =0$. The numerically computed levels seem to collapse to the single energy $E_{th}$ at $g_2=g_{\rm T}$, with a single bound state remaining below $E_{th}$. In reality, the spectrum is continuous above $E_{th}$ and the number of bound states is infinite.
(b) $A_4 =0.0001\omega$. The spectral collapse does no longer occur, because the effective potential is confining even though $A_4$ is very small.
(c) A zoom-in plot around $\omega=1.0\Omega$ of (b).
}
\label{fig-Ej-collapse}
\end{figure*}
\begin{figure*}[t]
\includegraphics[width=2\columnwidth]{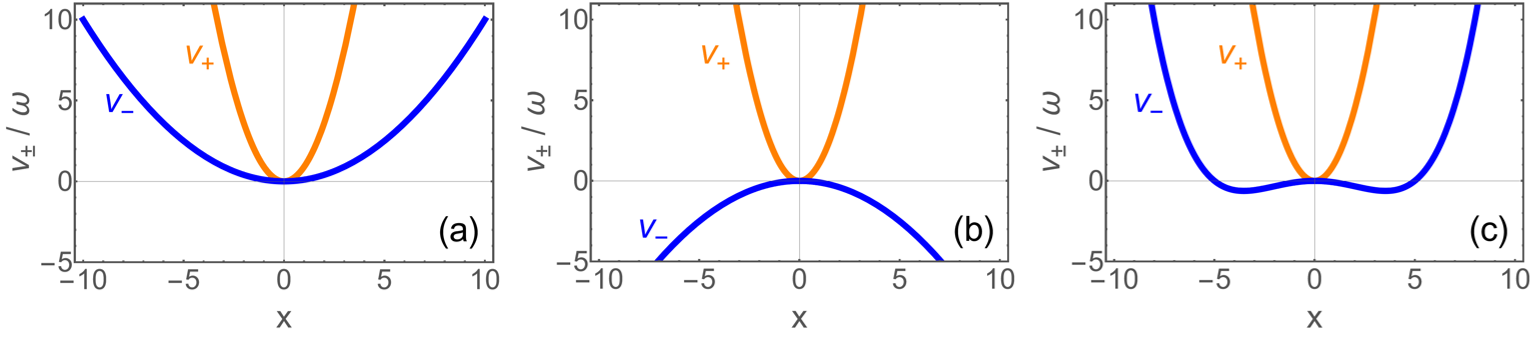}
\caption{Effective potential $v_{\pm}$ for $\Omega=0$:
(a) $A_4=0$, $g_2=0.8g_{\rm T}$.
(b) $A_4=0$, $g_2=1.2g_{\rm T}$.
(c) $A_4=0.001$, $g_2=1.2g_{\rm T}$.
}
\label{fig-v-potential}
\end{figure*}
\begin{figure*}[t]
\includegraphics[width=2\columnwidth]{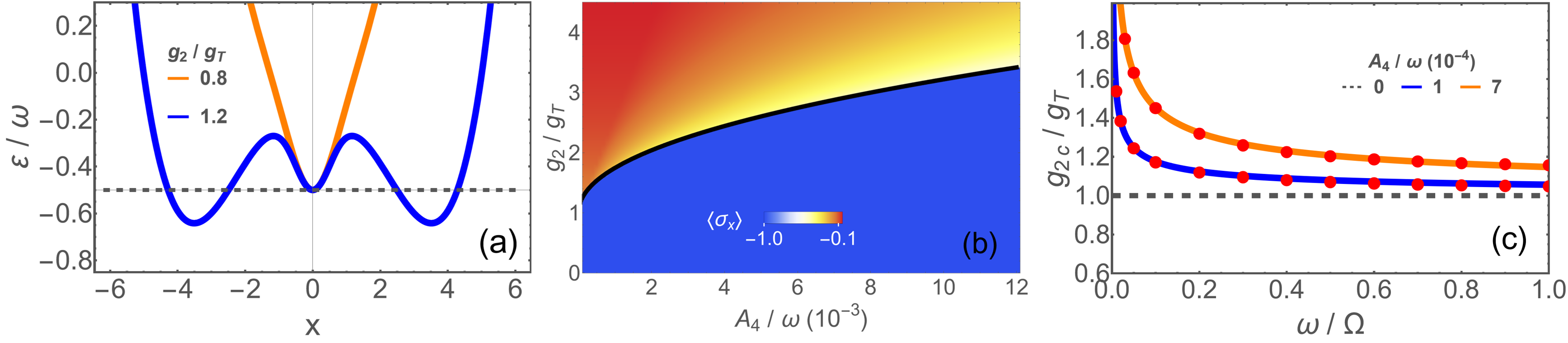}
\caption{Quantum phase transition.
(a) Variational energy $\varepsilon$ vs $x$ in the semiclassical limit $\omega/\Omega\rightarrow 0$ at $g_2=0.8g_T$ [orange (light gray)] and $g_2=1.2g_{\rm T}$ [blue (dark gray)].
(b) Phase diagram of $\langle \sigma_x \rangle$ (density plot) and analytic transition boundary $g_{2c}$ by Eq.\eqref{g2c} (black solid line) vs $A_4$ at $\omega=0.05\Omega$.
(c) $g_{2c}$ vs $\omega$ at $A_4=0.007\omega$ [orange (light gray)] and $A_4=0.001\omega$ [blue (dark gray)].
}
\label{fig-SemiClassical-e-gc}
\end{figure*}
\begin{figure*}[t]
\includegraphics[width=2\columnwidth]{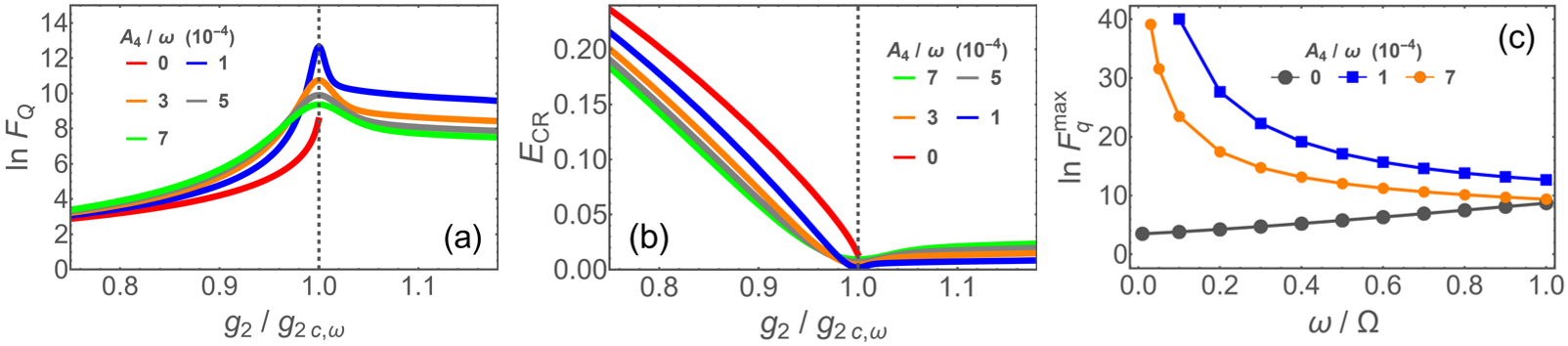}
\caption{Quantum Fisher information $F_Q$ for quantum metrology.
(a) $F_Q$ in natural logarithmic scale for different values of $A_4$ at $\omega=1.0\Omega$.
(b) The Cram\'{e}r-Rao error lower bound $E_{CR}$ corresponding to (a).
(c) Frequency dependence of the maximum values of $F_Q$ at different values of $A_4$. 
Here, $\lambda =g_2$ is taken as the measurement parameter for $F_Q$, $g_2$ is rescaled by the peak position $g_{2c,\omega}$ at frequency $\omega$,
and the unit scale of ratio $A_4/\omega$ is $10^{-4}$.
}
\label{fig-QFI-w}
\end{figure*}

\begin{figure*}[t]
\includegraphics[width=2\columnwidth]{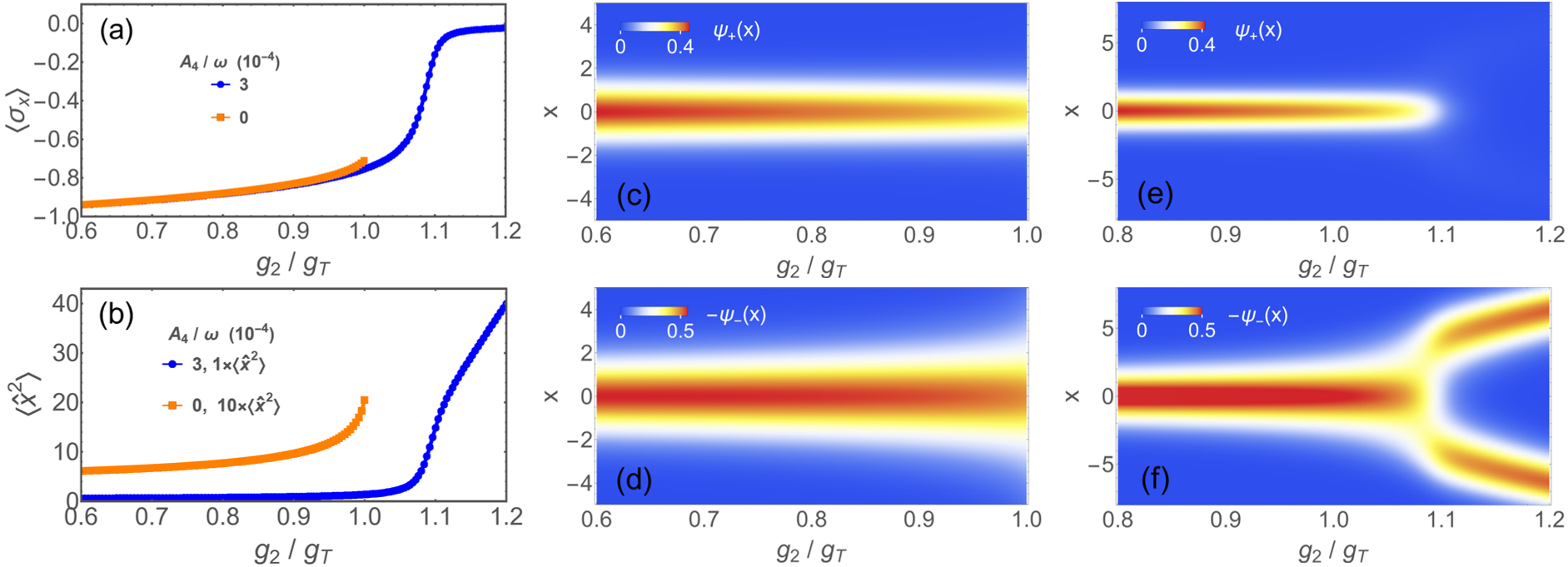}
\caption{Resource of sensitivity for critical quantum metrology.
(a,b) $\langle \hat{\sigma} _x\rangle$ (a) and $\langle \hat{x} ^2\rangle$ (b)
at $A_4=0$ [orange (light gray) squares] and $A_4=0.0003$ [blue (dark gray) dots]. In (b) the plot of $\langle \hat{x} ^2\rangle$ is amplified by 10 times for $A_4=0$.
(c-f) Evolution of the wave-function components $\psi _{+}(x)$ (c,e) and $\psi _{-}(x)$ (d,f) with respect to
$g_2$ at $A_4=0$ (c,d) and $A_4=0.0003$ (e,f).
Here, $\omega=1.0\Omega$ in all panels.
}
\label{fig-sens-resource}
\end{figure*}
\begin{figure*}[t]
\includegraphics[width=2\columnwidth]{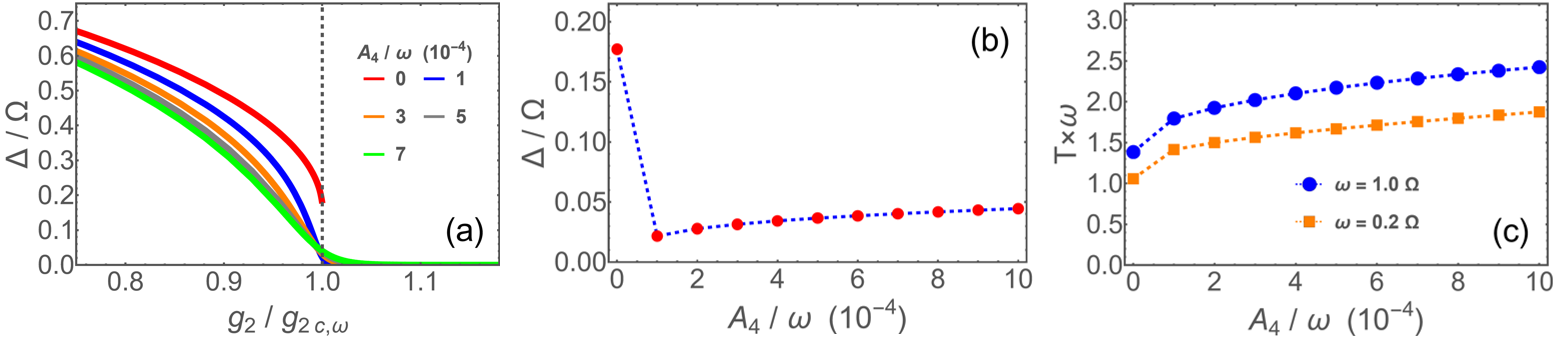}
\caption{Finite gap and preparation time for probe state (PTPS).
(a) Evolution of the gap $\Delta$ with respect to the coupling $g_2$ at different values of $A_4$.
(b) $\Delta$ at $g_{2c,\omega}$ versus $A_4$. Here, $\omega=1.0\Omega$ in (a) and (b).
(c) Dependence of the PTPS ($T$) on $A_4$ at different frequencies.
}
\label{fig-gap-Time}
\end{figure*}

\section{Absence of spectral collapse}
\label{Sect-Cure-ej-collapse}

The model (\ref{H-T}) with two-photon coupling exhibits the so-called ``spectral collapse'' phenomenon~\cite{Felicetti2015-TwoPhotonProcess,Felicetti2018-mixed-TPP-SPP,e-collpase-Lo-1998,e-collpase-Duan-2016,e-collpase-Garbe-2017,CongLei2019,Braak2023AnnPhys}.
This is illustrated in Fig. \ref{fig-Ej-collapse}(a), the energy levels of excited states $E_{i}$ seem to collapse to a single energy $E_{th}$ when the
coupling $g_{2}$ reaches the critical value $g_{{\rm T}}$. This apparent infinite degeneracy of $E_{th}$ is a numerical artifact, caused by the necessarily finite dimension of the state space used in exact diagonalization of the model. In reality, the spectrum exhibits a \emph{continuous} part, covering the whole real axis from  $E_{th}$ upward~\cite{e-collpase-Duan-2016,Braak2023AnnPhys}. Besides the continuum, the model \eqref{H-T} possesses  at $g_2=g_{\rm T}$ infinitely many bound states below $E_{th}$, which are exponentially close to $E_{th}$ and thus difficult to resolve by exact diagonalization~\cite{LiBraakChen2025}.

The origin of the spectral collapse can
be understood from a simple argument for the case $\Omega=0$. Going to the $\hat{x}/\hat{p}$-representation via
$a^{\dagger }=(\hat{x}-i\hat{p})/\sqrt{2},$ $a=(\hat{x}+i\hat{p})/\sqrt{2}$,
where $\hat{x}$ is the multiplication operator and $\hat{p}=-i\frac{\partial }{\partial x}$, we obtain~\cite{Irish2014,Ying2015,Ying-2018-arxiv,Ying2020-nonlinear-bias}
\begin{equation}
H_{\rm T}=\sum_{\sigma _{z}=\pm }h_{\sigma _{z}}\left\vert \sigma _{z}\right\rangle
\left\langle \sigma _{z}\right\vert +\frac{\Omega }{2}\sum_{\sigma _{z}=\pm
}\left\vert \sigma _{z}\right\rangle \left\langle \overline{\sigma }%
_{z}\right\vert   \label{Hx}
\end{equation}%
where $\sigma _{z}=-\overline{\sigma }_{z}=\pm $ labels the spin state in $z$
direction. Here $h_{\pm }=\frac{\omega }{2m_{\pm }}\hat{p}^{2}+v_{\pm
}\left( x\right) -\frac{1}{2}\omega $ is effective singe-particle
Hamiltonian, with effective mass $m_{\pm }=[1\mp \frac{(1-\chi )}{\left(
1+\chi \right) }\frac{g_{2}}{g_{{\rm T}}}]^{-1},$ in the spin-dependent
potential~\cite{CongLei2019,Ying-2018-arxiv,Ying2020-nonlinear-bias,Ying-g2hz-QFI-2024}
\begin{equation}
v_{\pm }(x)=\frac{1}{2}\omega \left( 1\pm \frac{g_{2}}{g_{{\rm T}}}\right)
x^{2}.  \label{v-A4=0}
\end{equation}
The spin-flip term proportional to $\Omega$ in \eqref{Hx} is relatively bounded with respect to the spin-diagonal term and plays no role in locating the critical point $g_{\rm T}$.
For $g_{2}<$ $g_{{\rm T}}$, the quadratic potential is positive for both spin values, as illustrated in Fig. \ref{fig-v-potential}(a), leading to a purely discrete spectrum. However, in the $g_{2}>g_{{\rm T}}$ regime, the
spin-down potential $v_{-}(x)$ is negative, as
illustrated in Fig.~\ref{fig-v-potential}(b), corresponding to an inverted harmonic oscillator (the parabolic barrier). The Hamiltonian is unbounded from below but still self-adjoint and generates a physically meaningful time evolution~\cite{ReedSimon1975}. The spectrum is purely continuous, spanning the whole real axis from $-\infty$ to $+\infty$, similar to the spectrum of $\hat{x}$ and $\hat{p}$.  At the boundary $g_{2}=g_{{\rm T}}$, the spin-down potential
$v_{-}(x)$ becomes flat, thus the spin-down sector exhibits a continuous spectrum for $\Omega=0$. Whether this continuum persists and/or is modified for non-zero $\Omega$, when both spin sectors are coupled, can be analyzed via the $G$-function technique~\cite{Braak2023AnnPhys,LiBraakChen2025}.

We have seen that the ``collapse'' is caused by an instability of the effective potential in the spin-down sector at $g_2=g_{\rm T}$. It can therefore be avoided by adding a positive potential which grows stronger than quadratically at infinity, namely the term
$A_4(a^{\dagger} +a)^4\sim \hat{x}^4$ in \eqref{H},
as demonstrated in Fig.~\ref{fig-Ej-collapse}(b)
where we have
a purely discrete spectrum for any value of $g_{2}$.
The effective potential for $\Omega=0$ reads now
\begin{equation}
v_{\pm }(x)=\frac{1}{2}\omega \left( 1\pm \frac{g_{2}}{g_{{\rm T}}}\right)
x^{2}+4A_{4}x^{4},
\end{equation}
with  $A_{4}>0$, as plotted in  Fig.~\ref{fig-v-potential}(c).

\section{Quantum phase transition in the slow-mode limit}
\label{Sect-QPT-A4}

A fascinating aspect of cavity QED is the presence of  quantum phase transitions due to the infinite-dimensional Hilbert space of a single light mode~\cite{
Liu2021AQT,Ashhab2013,Ying2015,Hwang2015PRL,Ying2020-nonlinear-bias,Ying-2021-AQT,LiuM2017PRL,Hwang2016PRL,Irish2017, Ying-gapped-top,Ying-Stark-top,Ying-Spin-Winding,Ying-2018-arxiv,Ying-JC-winding,Ying-Topo-JC-nonHermitian,Ying-Topo-JC-nonHermitian-Fisher,Ying-gC-by-QFI-2024,Ying-g2hz-QFI-2024, Grimaudo2022q2QPT,Grimaudo2023-Entropy,Zhu2024PRL}.
Here we find that the even in the absence of spectral collapse our non-linear system exhibits such a
quantum phase transition, which can be inferred from the effective potential discussed in section~\ref{Sect-Cure-ej-collapse}.

Indeed, in the $g_{2}<$ $g_{{\rm T}}$ regime the (classical) particle tends to reside around the origin $x=0$ which is the bottom of the potential as in Fig. \ref{fig-v-potential}(a),
while beyond $g_{{\rm T}}$ the two
potential minima  have $x\neq0$ corresponding to broken parity symmetry as in Fig. \ref{fig-v-potential}(c).
Such a transition occurs in the slow-mode limit $\omega\ll \Omega$. In this situation the wave packet becomes so narrow relatively to the
potential size that it can be regarded as a semiclassical particle with large mass
in an external field (the potential) while the quantumness of the two-level system is kept through the $2\times2$ matrix structure~\cite{Ying2020-nonlinear-bias}. In such a picture
the kinetic energy is neglected and the effective Hamiltonian reads
\begin{equation}
H_{x}=\left(
\begin{array}{cc}
e_{+}\left( x\right)  & \frac{1}{2}\Omega  \\
\frac{1}{2}\Omega  & e_{-}\left( x\right)
\end{array}%
\right) ,
\end{equation}%
where $e_{\pm }\left( x\right) =\frac{1}{2}\omega \left( 1\pm \frac{g_{2}}{%
g_{{\rm T}}}\right) x^{2}+4A_{4}x^{4}$, which has a lower energy branch
\begin{equation}
\varepsilon \left( x\right) =\frac{1}{2}\left[ \omega x^{2}+8A_{4}x^{4}-%
\sqrt{\Omega ^{2}+\left( \frac{g_{2}}{g_{{\rm T}}}\right) ^{2}\omega
^{2}x^{4}}\right] .
\end{equation}%
The energy minimum position $x_{\min }$ is determined by the minimization of
$\varepsilon \left( x\right) $. In Fig. \ref{fig-SemiClassical-e-gc}(a) we
plot two cases of $\varepsilon \left( x\right) $, where $\varepsilon \left(
x\right) $ has a minimum at the origin for a weaker coupling $g_{2}=0.8g_{%
{\rm T}}$ [orange (light gray)] while $x_{\min }$ is away from the origin for a
stronger coupling $g_{2}=1.2g_{{\rm T}}$ [blue (dark gray)]. Thus, we see
that upon increasing the coupling strength the transition occurs when the
energy at $x_{\min }$ is equal to the energy at the origin
\begin{equation}
\varepsilon \left( x_{\min }\right) =\varepsilon \left( 0\right) .
\label{Eq-Ex=E0}
\end{equation}%
Equation (\ref{Eq-Ex=E0}) leads to the critical coupling ratio $\overline{g}_{2c}$,
\begin{equation}
\overline{g}_{2c}\equiv\frac{g_{2c}}{g_{\rm T}}=\sqrt{\frac{2}{3}+\frac{1+432\alpha _{4}}{3f^{1/3}}+\frac{1}{3}%
f^{1/3}+16\alpha _{4}},  \eqnum{ }  \label{g2c}
\end{equation}%
where
\begin{equation}
f=1080\alpha _{4}-1+24\left( 972\alpha _{4}^{2}+\sqrt{6\alpha _{4}\left(
54\alpha _{4}-1\right) ^{3}}\right)
\end{equation}%
and $\alpha _{4}=A_{4}\Omega /\omega ^{2}.$

The validity of the analytic expression for $g_{2c}$ in (\ref{g2c}) is
demonstrated in Fig.~\ref{fig-SemiClassical-e-gc}(b) where the black solid
line, representing the analytic $g_{2c}$, matches the transition
boundary of the spin expectation $\langle \hat{\sigma}_{x}\rangle $
extracted numerically from the minimization of $\varepsilon \left( x\right) $.
Surprisingly, the validity of $g_{2c}$ extends to finite $\omega$, as
shown in Fig.~\ref{fig-SemiClassical-e-gc}(c) where the analytic $g_{2c}$ from \eqref{g2c} (solid lines)
agrees with the transition points defined by the peak of the QFI (dots) \cite{Ying-Topo-JC-nonHermitian-Fisher} which will be subject of section~\ref{Sect-higher-QFI-A4}.

We may expand $g_{2c}$ for  small $\alpha _{4}$ and large $\alpha
_{4}$, respectively,
\begin{eqnarray}
\overline{g}_{2c} &\approx &\sqrt{1+8\sqrt{2\alpha _{4}}+24\alpha _{4}},
\label{g2c-small-a} \\
\overline{g}_{2c} &\approx &\sqrt{\frac{2}{3}+16\alpha _{4}+12\alpha _{4}^{\frac{2}{3}%
}+4\alpha _{4}^{\frac{1}{3}}+\frac{\alpha _{4}^{-\frac{1}{3}}}{27}-\frac{%
\alpha _{4}^{-\frac{2}{3}}}{324}},  \label{g2c-large-a}
\end{eqnarray}%
the former for small $\alpha _{4}$ and the latter for large $\alpha _{4}
$. We see that in both regimes the expansions manifest an fractional power
law in $\alpha _{4}$. Such a fractional power law
in the dependence of the critical coupling on the frequency ratio is also found in the  linear quantum Rabi model \cite{Ying-gC-by-QFI-2024}. There the fractional power law comes from the frequency renormalization of
polarons, while here it stems from both nonlinear coupling terms, the quadratic and the quartic one.
A comparison of the approximated forms with the full expression for  $g_{2c}$ is
presented in Appendix \ref{Append-g2c-Expansion}.

\section{Quantum Fisher information for non-linear critical quantum metrology%
}\label{Sect-higher-QFI-A4}

In Fig.~\ref{fig-SemiClassical-e-gc}(c) we have compared the analytic expression \eqref{g2c} for $g_{2c}$ with the peak
position of the QFI which takes the following form for a pure state
$\psi \left( \lambda \right)$~\cite{Cramer-Rao-bound,Taddei2013FisherInfo,RamsPRX2018}
\begin{equation}
F_{Q}=4\left[ \langle \psi ^{\prime }\left( \lambda \right) |\psi ^{\prime
}\left( \lambda \right) \rangle -\left\vert \langle \psi ^{\prime }\left(
  \lambda \right) |\psi \left( \lambda \right) \rangle \right\vert ^{2}\right]
\label{QFI}
\end{equation}%
where $^{\prime }$ denotes the derivative with respect to the system parameter $\lambda $. In
quantum metrology the measurement precision of experimental estimation of
the parameter $\lambda $ is bounded from below by $F_{Q}^{1/2}$\cite{Cramer-Rao-bound},
with a higher QFI meaning a higher measurement precision. The QFI also
corresponds to the susceptibility of the fidelity $F$ via relation $\chi
_{F}=F_{Q}/4$\cite%
{Gu-FidelityQPT-2010,You-FidelityQPT-2007,You-FidelityQPT-2015},
\begin{equation}
F=\left\vert \langle \psi \left( \lambda \right) |\psi \left( \lambda
+\delta \lambda \right) \rangle \right\vert =1-\frac{\delta \lambda ^{2}}{2}%
\chi _{F},
\end{equation}
for an infinitesimal parameter variation $\delta \lambda $. Thus the
appearance of a peak in the QFI not only indicates the optimal condition for the metrological application of the system~\cite{Garbe2020,Garbe2021-Metrology,Ilias2022-Metrology,Ying2022-Metrology} but also the transition point itself~\cite{Zhou-FidelityQPT-2008,Zanardi-FidelityQPT-2006,Gu-FidelityQPT-2010,You-FidelityQPT-2007,You-FidelityQPT-2015},
as demonstrated in~\cite{Ying-gC-by-QFI-2024}. We note here that the
second term in \eqref{QFI} vanishes and the expression for $F_Q$ simplifies as
\begin{equation}
F_{Q}=4\langle \psi ^{\prime }\left( \lambda \right) |\psi ^{\prime }\left(
\lambda \right) \rangle
\end{equation}%
for a real wave function $\vert\psi(\lambda)\rangle$, which describes non-degenerate states of real Hamiltonians~\cite{Ying-gC-by-QFI-2024}.

Figure \ref{fig-QFI-w}(a) shows the QFI around the transition for
different values of $A_{4}$ and finite frequency $\omega =\Omega $. The parameter $\lambda$ to be measured is $\lambda=g_{2}$. In the figure, the coupling strength $g_{2}$ is
rescaled by the peak position $g_{2c,\omega}$. The
QFI becomes exponentially large around $g_{2c,\omega}$ (the $y$-axis in Fig.~\ref{fig-QFI-w}(a) has a logarithmic scale), meaning very high precision. The
Cram\'{e}r-Rao lower bound for the error \cite{Cramer-Rao-bound}, $E_{CR}=F_{Q}^{-1/2}$, is small around $g_{2c,\omega }$ as shown in Fig. \ref{fig-QFI-w}(b).
Remarkably,
adding a small $A_{4}$-term leads to a higher peak value of the  measurement precision than the case without regulator, $A_{4}=0$. Furthermore, for $A_{4}\neq 0$ (but sufficiently small),  $F_{Q}$ diverges for $\omega\rightarrow 0$, while $F_{Q}$ for $A_{4}=0$  decreases, as seen in Fig.~\ref{fig-QFI-w}(c). This shows that adding a small
quartic term can dramatically improve the measurement
precision in quantum metrology.

\section{Mechanism for the enhanced precision}
\label{Sect-Mechanism-A4}

We may clarify the mechanism behind the much higher precision
reached by adding a small $A_{4}$ term resorting to the variation of the ground state wave function with the parameter $\lambda=g_2$. We compare the expectation values of $\langle \sigma _{x}\rangle $
and $\langle \hat{x}^{2}\rangle $ in Figs.~\ref{fig-sens-resource}(a) and
\ref{fig-sens-resource}(b) at finite frequency $\omega =1.0\Omega $ as functions of $g_2$.
For $A_{4}=0$, both $\langle \sigma _{x}\rangle $ and $\langle \hat{x}^{2}\rangle$ show a cusp in the vicinity of
the spectral collapse point $g_{2}=g_{{\rm T}}$, clearly visible in the
amplified plot of $\langle \hat{x}^{2}\rangle $ in Fig.\ref{fig-sens-resource}(b). For $g_2>g_{{\rm T}}$, the system becomes unstable and has no longer a ground state.
In contrast, for $A_4\neq 0$, $\langle \sigma_{x}\rangle$ and $\langle \hat{x}^{2}\rangle$ show critical behavior with a sign change of the their second derivative at a coupling value $g_{2c}$ slightly larger than $g_{{\rm T}}$, but no instability.

The sensitivity difference is rooted in the distinct pattern of variation of the wave
function with $g_2$, as it enters directly the formula for the QFI in \eqref{QFI}. Figs.\ref{fig-sens-resource}(c)-\ref{fig-sens-resource}(f) show
the evolution of the two components $\psi_\pm(x)$ of the ground state wave function with respect to the coupling
$g_{2}$ in the absence [Figs.\ref{fig-sens-resource}(c) and \ref%
{fig-sens-resource}(d)] and presence [Figs.\ref{fig-sens-resource}(e) and %
  \ref{fig-sens-resource}(f)] of the $A_{4}$-term.
We see that for $A_4=0$, the wave function $\psi _{+}$ ($\psi _{-}$) is slightly narrowing (broadening), while approaching the
spectral collapse point $g_{2}=g_{{\rm T}}$. Such wave packet narrowing and
broadening are driven by the potential narrowing and broadening due to the
frequency renormalization factor $(1\pm g_{2}/g_{{\rm T}})$ in Equ.~(\ref{v-A4=0}), as also plotted in Fig.~\ref{fig-v-potential}(a) and are also a consequence of the fact that $\psi_+$ and $\psi_-$ are mutual Fourier transforms of each other for $g_2=g_{{\rm T}}$~\cite{Braak2023AnnPhys}. In contrast,
the wave-function variation in the presence of the $A_{4}$ term is
coming not only from the frequency renormalization but also from the wave-packet
bifurcating and rapid shifting away from the origin, as in
Fig. \ref{fig-v-potential}(f). The wave-packet shifting leads to a faster
change with $g_2$ than the mere narrowing and broadening, thus entailing a larger value of
$\langle \psi ^{\prime }\left( \lambda \right) |\psi ^{\prime }\left(\lambda \right) \rangle$ and therefore a
higher sensitivity. As a result, a much higher measurement precision is
reached for $A_4\neq 0$.

\section{Finite gap and preparation time for the probe state}
\label{Sect-ProbeTime-A4}

Besides the measurement precision characterized by the QFI another aspect
important for the practical implementation of quantum metrology is the preparation time for the probe state
(PTPS)~\cite{Garbe2020,Ying2022-Metrology,Ying-g2hz-QFI-2024,Gietka2022-ProbeTime}. The PTPS in critical quantum
metrology is determined by the gap between ground state and first excited state, with longer PTPS needed for smaller
gaps. In the linear QRM the PTPS diverges at the critical point due to the exponentially fast closing of the gap in the slow-mode limit~\cite{Garbe2020,Ying2022-Metrology,Ying-g2hz-QFI-2024}. This problem can be remedied by using
a mixed model with both linear and quadratic light-matter interactions present~\cite{Ying2022-Metrology}, leading to a finit PTPS at the critical point.

Here we find that the PTPS is finite both in the absence and presence of the
$A_{4}$ term. We present the gap evolution with respect to the coupling
$g_{2}$ for different strengths of $A_{4}$ in Figs.~\ref{fig-gap-Time}(a,b). We
see that the gap is finite in all cases at $g_{2c,\omega }$ where the QFI reaches
the maximum. The PTPS  can be estimated as~\cite{Garbe2020,Ying2022-Metrology}
\begin{equation}
T=\int_{0}^{1}\frac{1}{\Delta \left( \overline{g}_{2}\right) }d\overline{g}%
_{2}
\end{equation}%
where $\overline{g}_{2}=g_{2}/g_{2c,\omega }$. Despite the larger gap for $A_{4}=0$, the PTPS stays always finite and has
the same order for zero and finite $A_{4}$, as seen in Fig.\ref{fig-gap-Time}(c). Note here that, although the PTPS is inversely
proportional to the frequency, the PTPS is still finite in our case because the collapse instability from the quadratic term provides a critically enhanced sensitivity resource without requiring the limit $\omega\rightarrow 0$~\cite{Ying2022-Metrology}.
Remarkably, this feature persists upon adding the $A_{4}$-term which removes the instability and even leads to a larger QFI as compared to the conventional two-photon QRM. One may thus realize much higher
orders of measurement precision without paying the price of a
diverging PTPS.

\begin{figure}[t]
\includegraphics[width=0.75\columnwidth]{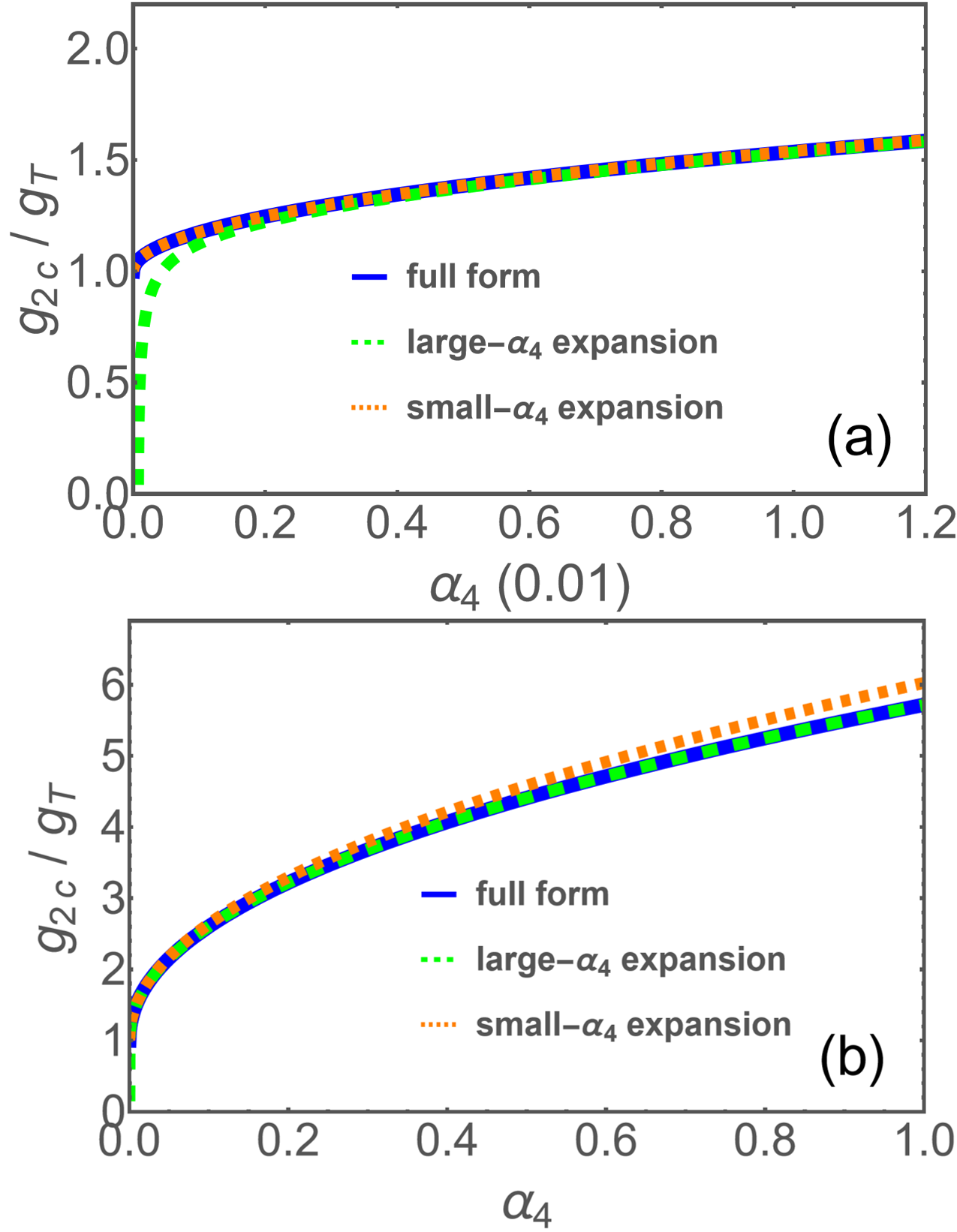}
\caption{Comparison of $g_{2c}$ expressions in the
full form [Eq. (\ref{g2c}), blue solid line] and the expansion forms by small-$\alpha _{4}$ expansion [Eq. (\ref{g2c-small-a}), orange dotted line)] and
large-$\alpha _{4}$ expansion [Eq. (\ref{g2c-large-a}), green dashed line]. (a)
$\alpha _{4}\in  [0,0.012]$ (b) $\alpha _{4}\in [0,1]$.
}
\label{fig-g2c-Expansion}
\end{figure}

\section{Conclusions}
\label{Sect-Conclusions-A4}

In this work we have studied a quadratically coupled QRM with an additional regulator term, quartic in the boson operators, with focus on its application to
non-linear critical quantum metrology.
The ``spectral collapse'' phenomenon of
the standard two-photon QRM does no longer occur and the spectrum stays bounded from below and purely discrete for all parameter values.
The introduction
of the quartic $A_{4}$ term induces a quantum phase transition in the low-frequency
limit, for which we have obtained the phase boundary line analytically.
The formula shows a fractional power law in the dependence of the coupling
strength. The phase transition extended for finite frequency ratio can be employed for non-linear critically enhanced quantum metrology. Indeed, by examining
the QFI we find that adding the $A_{4}$ term can yield a much higher order
of measurement precision than the conventional two-photon QRM. This enhanced precision is shown to be caused by the different pattern of variation of the ground state wave function while tuning through the
transition. Finally we have checked the excitation gap and the PTPS. It turns
out that the PTPS is finite both in the absence or presence of the $A_{4}$
term. Therefore, adding the $A_{4}$ term leads to a metrology protocol which exhibits a strongly enhanced measurement precision compared to the conventional two-photon QRM, while keeping the same \emph{finite} value of the PTPS. As a final
remark we would like to mention that the proposed quadratic QRM with the
additional quartic term can be realized in superconducting circuit
systems using current technology.

\section*{Acknowledgments}

This work was supported by the National Natural Science Foundation of China
(Grants No. 12474358, No. 11974151, and No. 12247101). S.F. acknowledges financial
support from PNRR MUR project PE0000023-NQSTI financed by the European Union - Next Generation EU.
D.B. acknowledges support from the German Research Foundation (DFG) under grant No. 439943572.

\bigskip
\appendix

\section{Comparison of $g_{2c}$ expressions in full form and expansion forms}
\label{Append-g2c-Expansion}

In Fig. \ref{fig-g2c-Expansion}, we compare the $g_{2c}$ expressions in the
full form [Eq. (\ref{g2c}), blue solid line] and the expansion forms by
small-$\alpha _{4}$ expansion [Eq. (\ref{g2c-small-a}), orange dotted line)]
and large-$\alpha _{4}$ expansion [Eq. (\ref{g2c-large-a}), green dashed
line]. Indeed, for the $\alpha _{4}\in \lbrack 0,0.01]$ regime in Fig. \ref%
{fig-g2c-Expansion}(a) we see that expansion (\ref{g2c-small-a}) reproduces
the full-form result, while expansion (\ref{g2c-large-a}) has some
qualitative deviations in the small-$\alpha _{4}$ limit. For the $\alpha
_{4}\in \lbrack 0.01,\infty ]$ regime illustrated in Fig. \ref%
{fig-g2c-Expansion}(b), expansion (\ref{g2c-large-a}) works very well for
the entire regime, while expansion (\ref{g2c-small-a}) has some considerable
deviations.

\bibliography{Refs-tex-nonlinear--g2-A4--dropbox-Merge-3-arXiv}

\end{document}